# Leg-tracking and automated behavioral classification in *Drosophila*

Jamey Kain[1], Chris Stokes[1], Quentin Gaudry[2], Xiangzhi Song[1,3], James Foley[1], Rachel Wilson[2], Benjamin de Bivort[1,4,5]

[1]The Rowland Institute at Harvard, Cambridge, Massachusetts, USA. [2]Department of Neurobiology, Harvard Medical School, Boston, Massachusetts, USA. [3]College of Chemistry & Chemical Engineering, Central South University, Changsha, P. R. China. [4]Center for Brain Science, Harvard University, Cambridge. Massachusetts, USA.
[5]Department of Organismic and Evolutionary Biology, Harvard University, Cambridge Massachusetts, USA.

Correspondence should be addressed to B.D. (debivort@rowland.harvard.edu).

Here we present the first method for tracking each leg of a fruit fly behaving spontaneously upon a trackball, in real time. Legs were tracked with infrared-fluorescent dye invisible to the fly, and compatible with two-photon microscopy and controlled visual stimuli. We developed machine learning classifiers to identify instances of numerous behavioral features (e.g. walking, turning, grooming) thus producing the highest resolution ethological profiles for individual flies.

A major goal of biology is to elucidate the mechanisms underlying behavior and how they have evolved. Due to its vast genetic toolkit, *Drosophila melanogaster* is an ideal model system for understanding the underpinnings of behavior. However, analyzing behavior is not trivial and therefore most studies focus on simple, robust behaviors such as phototaxis [1] and olfactory chemotaxis [2]. Classic paradigms, performed on groups of flies, quantify these behaviors efficiently, but coarsely. In contrast, methods to efficiently characterize the behavior of individual flies at high levels of detail are rare. With recent advances in computer-vision, a new generation of automated and sophisticated assays are being developed that allow for richer characterizations of behaviors, and consequently the genes, circuits and evolution underlying them.

Several previous studies have introduced such sophisticated methods for the extraction of high-resolution behavioral data [3-9], but all these approaches have at least one of the following limitations: 1) They do not allow simultaneous monitoring of neural activity, 2) they require intensive manual analysis, 3) they only capture short bouts of activity, 4) they can only automatically detect a limited repertoire of behaviors, and 5) they can only track the legs of large insects like cockroaches.

Here we present a method that has none of these limitations. It is the first technique that tracks the six individual legs and the fictive movement of a tethered fruit fly, in real time, behaving either spontaneously or in response to controlled visual stimuli, for hours. We also developed a machine learning classifier that automatically detects and categorizes distinct behavioral features (e.g. walking, turning, grooming), rapidly providing the most detailed behavioral recordings of single flies yet achieved. Notably, this setup was designed for compatibility with either electrophysiology or optophysiological two-photon microscopy. Finally, we demonstrate that this novel methodology reveals new insights about *Drosophila* behavior. Specifically, we discovered significant individual-to-individual variation in freely behaving animals, and we found that individual variation is amplified by breaking the loop between motor behavior and sensory feedback.

We started with a traditional floating ball treadmill rig [7-9] since tethering a behaving fly simplifies leg imaging and permits the future incorporation of simultaneous electrophysiology or optophysiology. We substituted a transparent ball to allow imaging of the legs from below using a custom imaging system (Fig. 1a, Supplementary Fig. S1, Supplementary Note 1). The sphere was tracked by two infrared laser sensors normally found in computer mice. Since each sensor only detects two dimensions of motion, two sensors were used to capture all three rotational components (forward/backward, turning in place, and sidestep/crabwalk) (Supplementary Fig. S1d-f, Supplementary Note 2).

Small pieces (~100x100x50μm) of the dyes [2.2.1]-oxazine (221ox) [10] and julolidine-oxazine (julox) dye [11] were glued in alternation to each of the six legs (Fig. 1b, c, Supplementary Note 3). Flies were allowed to adapt to the dye spots for at least 24 hours, and then mounted above the floating sphere via a wire tether glued to the thorax. A HeNe laser illuminated the dye on the legs, which were imaged from below, through the clear sphere, by two cameras (Fig. 1a). Each camera was equipped with a band-pass filter optimized for either 221ox or julox, allowing each leg to be uniquely detected, and adjacent legs distinguished (Fig. 1d, e). The optical wavelengths used here were chosen for their invisibility to the fly (between 630 to 850 nm) [12] and compatibility with two-photon and traditional fluorescent microscopy (Fig. 1e).

Animals were typically recorded in the dark for 2 hours, although we have observed flies to behave beyond 16 hours. After data acquisition, the fly can easily be removed from the tether and saved for future use. Custom LabView software records 15 vectors in real-time: the $x$ and $y$ coordinates of each of the six legs, and the three rotational components of the floating sphere. Representative data is shown in Fig. 1f (see Supplementary Fig. S1g and Supplementary Note 5), including instances of grooming and walking behaviors. In addition to directly observing single leg strides via the dye and optics, the sphere itself was also responsive to single leg strides (Supplementary Figure S1h).

Due to the volume of data collected (15 vectors per frame at 80+Hz), we developed an automatic method for scoring behavior using non-linear classifiers, specifically $k$-nearest neighbors analysis (Fig. 2a, Supplementary Note 4) [13]. In order to train the classifier, two investigators independently scored, frame-by frame, movies recorded concurrently with leg tracking (Fig. 2b) (this training data set only needs to be generated once and can be used to score all future trials). Hand-scoring consisted of assigning to each frame of the movie one of 12 possible behavioral labels (Fig. 2c). The two investigators assigned the same score to 71% of frames, setting the goal for classifier performance (Fig. 2c). Classifiers trained on the original 15 vectors (the raw data) were mediocre. However, augmenting the raw data with higher order features, specifically the derivatives and local standard deviations of each of the 15 raw data vectors substantially improved performance. Lastly, applying a low-pass filter to the classifier predictions further improved their accuracy. In the end, the classifier had an error rate only 5% greater than that seen between independent manual scoring (Fig. 2d-f) (Supplemental Movie S1).

With the classifier in hand, we categorized the behavior of individual flies (e.g. Supplemental Movie S2) across 2 hour trials. Flies spent most of their time pausing or grooming and ran more at the start of trials (Fig. 3a, b). We built ethograms [14] to look at the direction and frequency of transitions between the 12 behavioral types (Supplementary Note 5). We found that the most frequent transitions were between the two types of foreleg grooms and the two types of hind leg grooms (Fig. 3d). However, individual flies also showed variation – e.g., one animal transitioned frequently from abdomen grooming to hind leg grooming, whereas another animal did so more rarely. By contrast, the latter animal frequently alternated between abdomen grooming and pausing, while the former did so rarely (Fig. 3d).

This individual-to-individual variation could be an artifact of mounting bias. To examine this, we remounted the same individual flies on subsequent days. We found that the transition rates in the ethograms were significantly more similar between re-mountings of the same fly than between flies, suggesting that much of the inter-trial variability comes from idiosyncrasies of the flies that persist for at least a few days (Fig. 3e). This was not observed (with statistical significance) in the simple percentage of time spent doing each behavior (Fig. 3c). Moreover, the idiosyncratic nature of behavioral transitions is evident in fine-scale analysis of the mean positions of individual legs across said transitions (Fig. 3f). Thus, we conclude that certain ethological characterizations, namely the transitions between behaviors, are robust to any mounting artifacts, persist for days, and vary across flies.

Since the setup uses wavelengths invisible to the fly (Fig. 1e), we can present controlled visual stimuli to the animal to study behaviors such as phototaxis or optomotor responses [15]. As a demonstration, we recorded animals' responses to optic flow by placing them in front of an LCD monitor displaying vertical bars translating from a line of expansion under either open- or closed-loop feedback control. We found that flies showed less variance in their average running speed during closed-loop trials relative to open-loop ($p=0.002$, $\chi^2$ test of variance Fig. 3g), an effect also seen in "flight-simulator" assays (pers. comm. Michael Reiser). The percentage of time spent doing each behavior accounts for the overall differences in activity we observed across the modulations of visual feedback (Fig. 3h).

This method provides unprecedented level of detail for the characterization of walking behavior, revealing rich and diverse behavioral profiles of individual animals. We also found that under closed-loop conditions, the flies behave more similarly to each other; perhaps interactive stimuli engage circuitry that overrides their more idiosyncratic preferences. The automatic classification of behavior described here represents a general approach to developing automatic ethological classifiers for the efficient collection of statistically powerful data sets. This method will be a powerful tool for descriptive and comparative studies, for analyzing subtle mutant phenotypes, and upon integration with optophysiological recording, for probing the activity of neural circuits as they mediate individual behaviors.

**METHODS**
See Supplemental Notes below.

**ACKNOWLEDGEMENTS**
We thank D. Cox and N. Pinto for guidance with non-linear classifiers, and S. Buchanan for comments on the manuscript. J. Kain, C. Stokes, X. Song, J. Foley and B. de Bivort were supported by the Rowland Institute. Q. Gaudry was funded by a research project grant from the

**AUTHOR CONTRIBUTIONS**
J.K. and B.D. conceived the study, fabricated the apparatus and performed the experiments. Q.G. and R.W. helped with early versions of the spherical treadmill. C.S. helped fabricate the apparatus. X.F. and J.F. synthesized the dyes. All authors helped prepare the manuscript.

**COMPETING FINANCIAL INTERESTS**
The authors declare no competing financial interests.

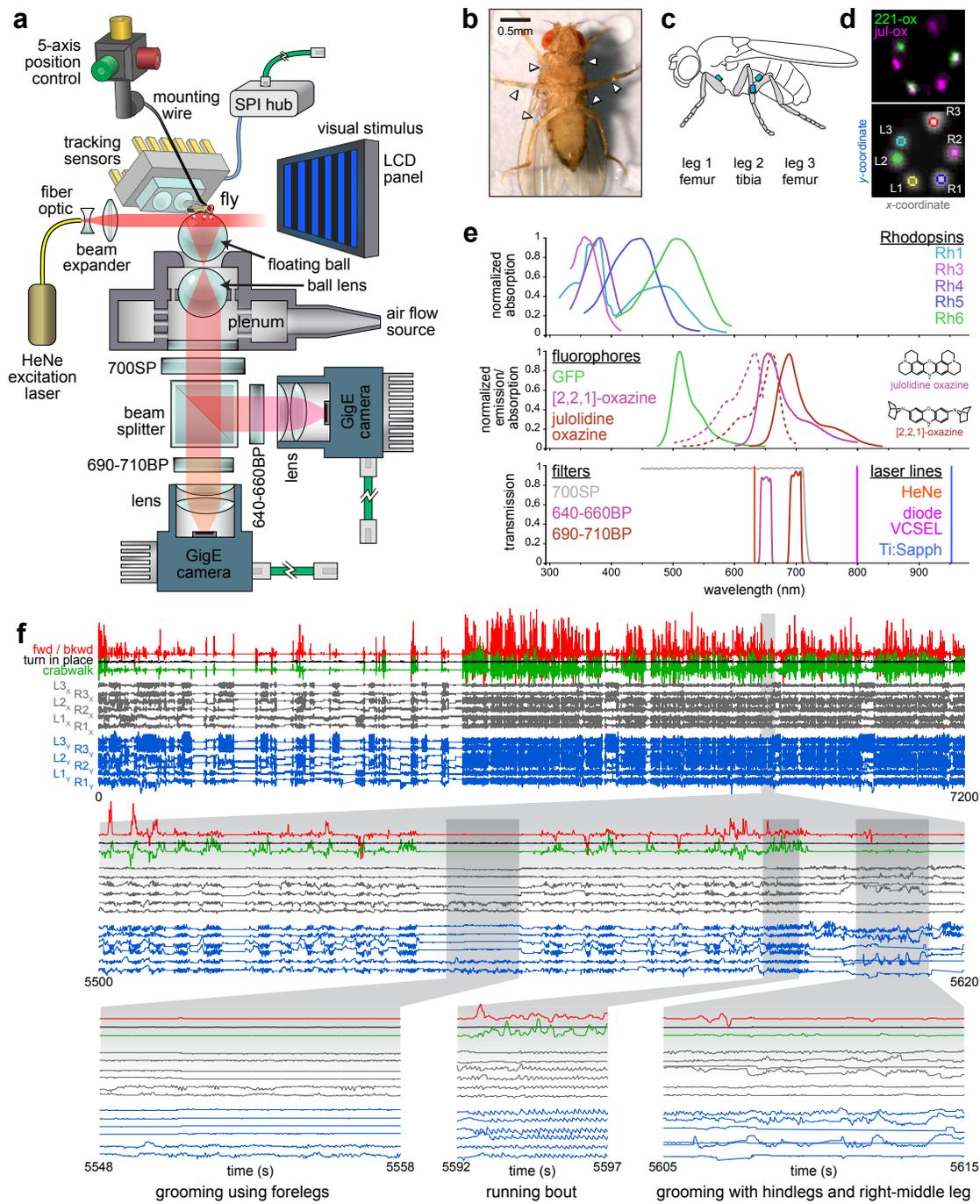

**Figure 1** | The leg tracker apparatus and its properties.
(a) Schematic of the leg tracker apparatus. A fly is mounted by wire tether to a clear sphere supported by flowing air. Fictive motion of the fly is recorded by tracking sensors. The dye spots on the legs fluoresce when excited by the HeNe laser and a second sphere is used to re-collimate the image of the dye spots which are tracked from below by two cameras.
(b) Photograph of a fly with bits of 221ox and julox dyes (arrowheads) glued alternately on each leg.
(c) Cartoon illustrating the positioning of the dye spots. Dye was placed upon the femurs of the fore and hind limbs and upon the tibia of the middle legs.
(d) Imaging of dye spots and detection of their positions. Top panel shows differential detection of alternating legs marked with the two dyes. Since the dyed segments on the front and hind legs do not cross, leg identity can be readily inferred in single frames (bottom panel).
(e) Diagram of the absorption wavelengths of the *Drosophila* rhodopsins (top). Diagram of the absorption (dashed lines) and emission (solid lines) wavelengths of the 221ox and julox dyes and their chemical structures (middle). The dyes are compatible with GFP. Emission wavelengths of the HeNe laser, vertical cavity surface emitting laser (VCSEL) of the tracking sensor and Ti:Sapphire lasers are depicted as vertical lines (bottom). Also shown are the transmission properties of the rig dichroic filters.
(f) Representative data showing the 15 data vectors (3 for ball motion, and 12 for leg positions) being recorded over 7200 seconds (top). Magnification of a 120 second data slice (middle). Higher magnifications of three typical behaviors (bottom). Red, black and green traces represent forward/backward, turning in place, and crabwalk motion respectively. Gray and blue traces are the *x* and *y* coordinates, respectively, of each leg. Plots are in arbitrary units, but actual measurements are calibrated.

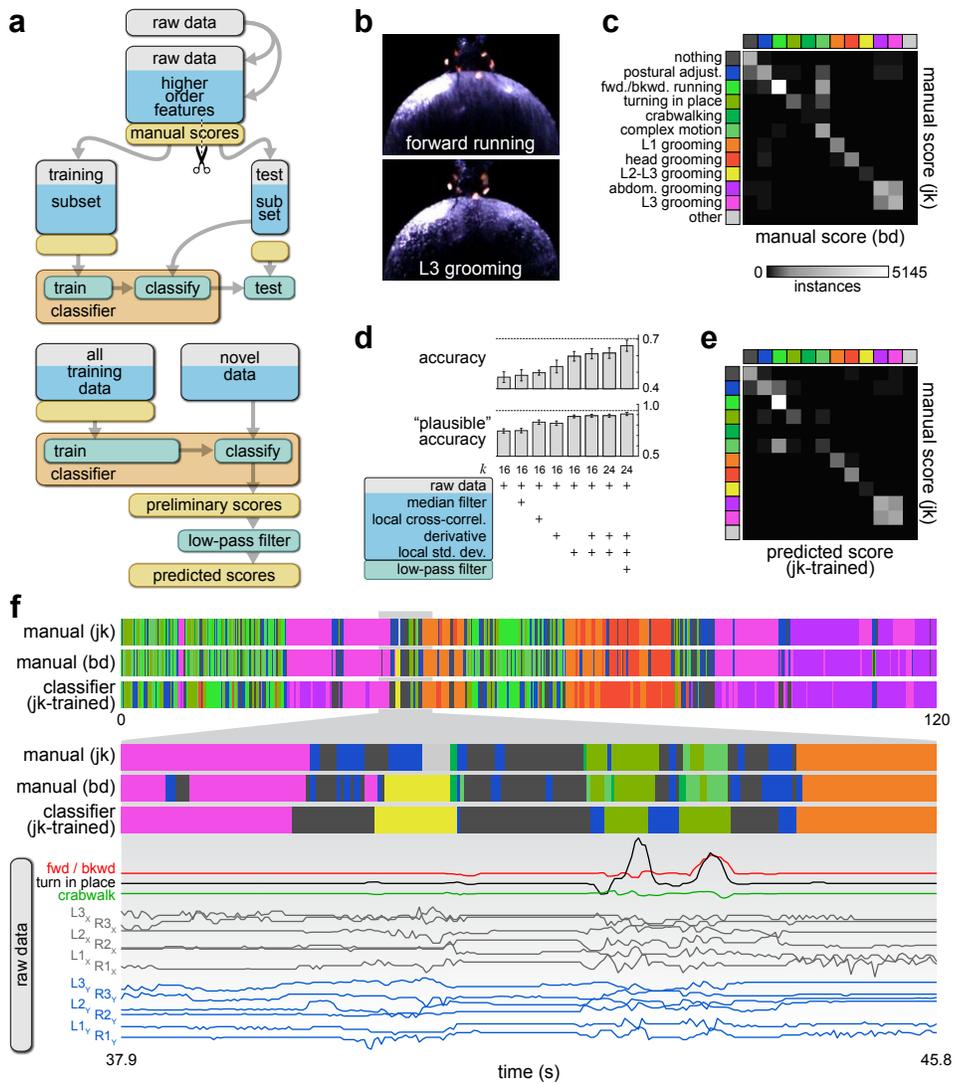

**Figure 2** | *k*-nearest neighbors classifier for the automatic labeling of data.
(a) Flowchart depicting development of the classifier (top) and its use to classify data (bottom) (see also Supplementary Note 4).
(b) Examples of movie frames used to hand-score behavior for training the classifier.
(c) Contingency matrix of the hand-scored movie-frame data sets from the two researchers. Shade of gray indicates the abundances of (dis)agreement between the two researchers. Colors here and in (e) and (f) indicate behavior types.
(d) Including higher order features with the raw data increases the classifier's accuracy and "plausible" accuracy (see Supplementary Note 4).
(e) Contingency matrix of the JK hand-scored data set to the classifier trained on the JK-scored data set.
(f) Sequences of behavior scores from both JK and BD manual data sets and from the classifier for the same 120sec window (top). Magnification of a ~8sec subset along with the raw data used by the classifier (bottom).

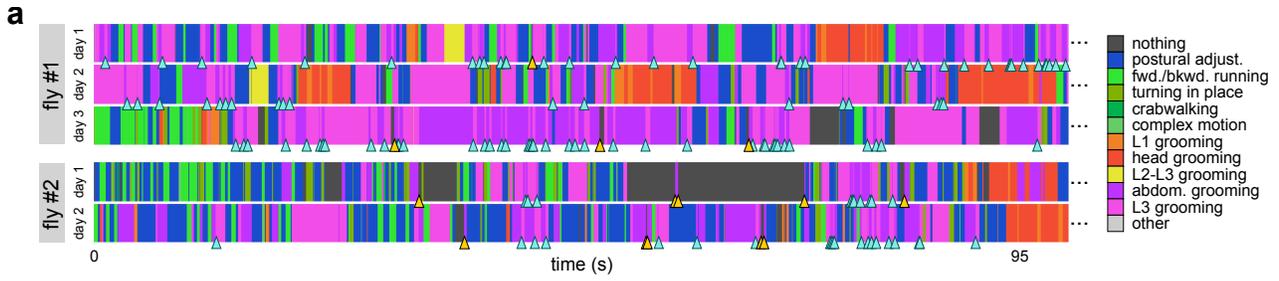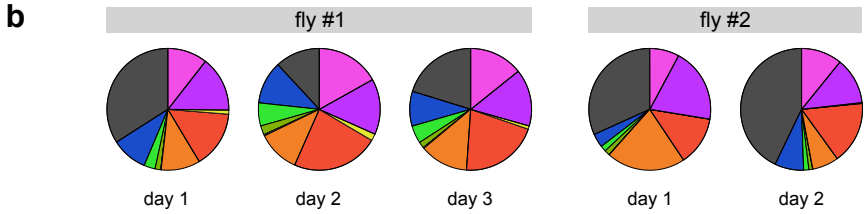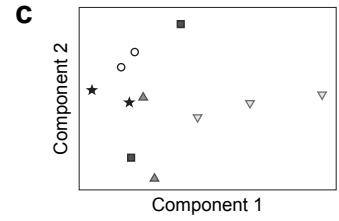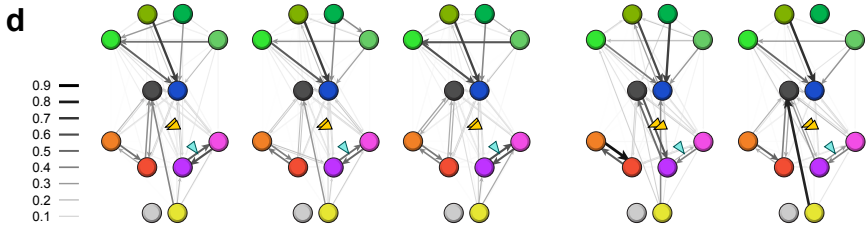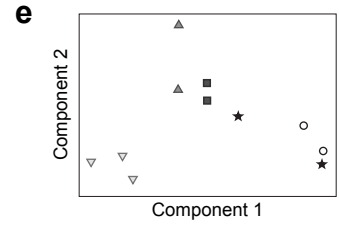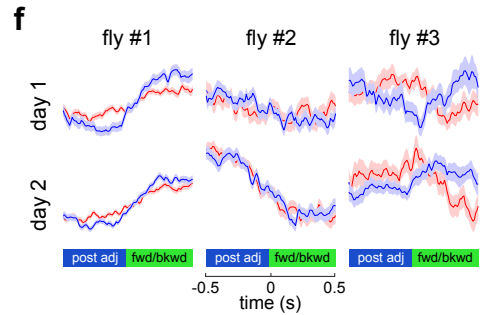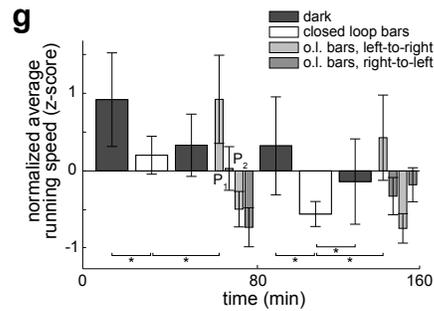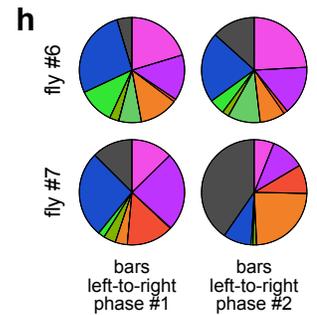

**Figure 3** | The behavioral patterns of individual animals.
(a) Multi-color bars show snapshots (95sec) of classifier-scores for multiple trials of two individual flies. Blue triangles indicate a behavioral transition more frequent in fly #1 and yellow triangles a transition more frequent in fly #2. Colors here and in (b), (d) and (h) indicate behavior type. In (a-f) distinct trials for each fly were recorded on separate days.
(b) Pie charts depicting fraction of time spent performing each behavior for distinct trials from two individual animals.
(c) Spatial visualization of principal components 1 and 2 underlying the inter-trial variance in fraction of time spent per behavior in (b). Shapes indicate multiple trials of an individual fly. See Supplementary Note 5.
(d) Ethograms of the same trials from (b). Arrows indicate direction of transition and thickness indicates frequency. Self-transitions are omitted for clarity. Triangles as in (a). See Supplementary Note 5.
(e) Spatial visualization of principal components 1 and 2 underlying the variance in the ethograms from (d). Shapes indicate multiple trials across days of an individual fly. See Supplementary Note 5.
(f) Individuality in the average *y*-coordinates of the left (red) and right (blue) hind legs during the transition from postural adjustment to motion. Shaded regions indicate SEM.
(g) Average *z*-score normalized running speed during optomotor experiments as a function of stimulus. Inter-animal variance decreases during closed-loop trials relative to open-loop and darkness ($p<0.0005$, $n=5$, $\chi^2$ test of variance). o.l. indicates open-loop phases. $P_1$ and $P_2$ indicate the phases analyzed in (h). Asterisks indicate significant differences in inter-animal variance between groups. Error bars are SEMs. See Supplementary Note 1, 5.
(h) Pie charts showing fraction of time spent performing each behavior during the first and second left-to-right moving bars open-loop phases, for two different flies. The diminishment in mean running speed during open-loop conditions (g) is reflected in increased stasis and diminished forward running in phase 2.

Supplemental Information for:

**Leg-tracking and automated behavioral classification in *Drosophila***

Jamey Kain[1], Chris Stokes[1], Quentin Gaudry[2], Xiangzhi Song[3], James Foley[1], Rachel Wilson[2], Benjamin de Bivort[1,4,5]

[1]The Rowland Institute at Harvard, Cambridge, Massachusetts, USA. [2]Department of Neurobiology, Harvard Medical School, Boston, Massachusetts, USA. [3]College of Chemistry & Chemical Engineering, Central South University, Changsha, P. R. China. [4]Center for Brain Science, Harvard University, Cambridge. Massachusetts, USA.
[5]Department of Organismic and Evolutionary Biology, Harvard University, Cambridge Massachusetts, USA.

Correspondence should be addressed to B.D. (debivort@rowland.harvard.edu).

Contents:
Supplemental Data
    Figure S1
    Movie S1
        http://www.youtube.com/watch?v=oP3cQ09zZu8
    Movie S2
        http://www.youtube.com/watch?v=miMGQHsAPQo
Supplemental Notes
    Note S1
    Note S2
    Note S3
    Note S4
    Note S5
Supplemental References

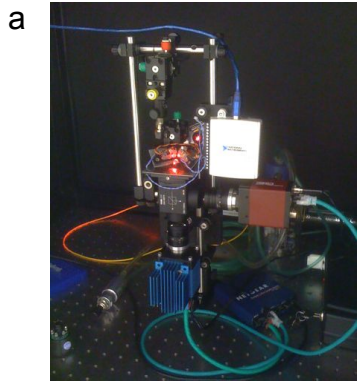
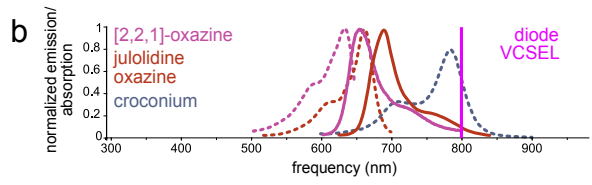
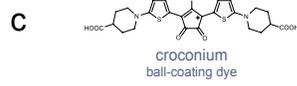
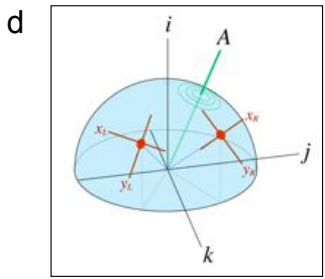
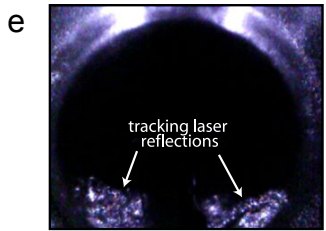
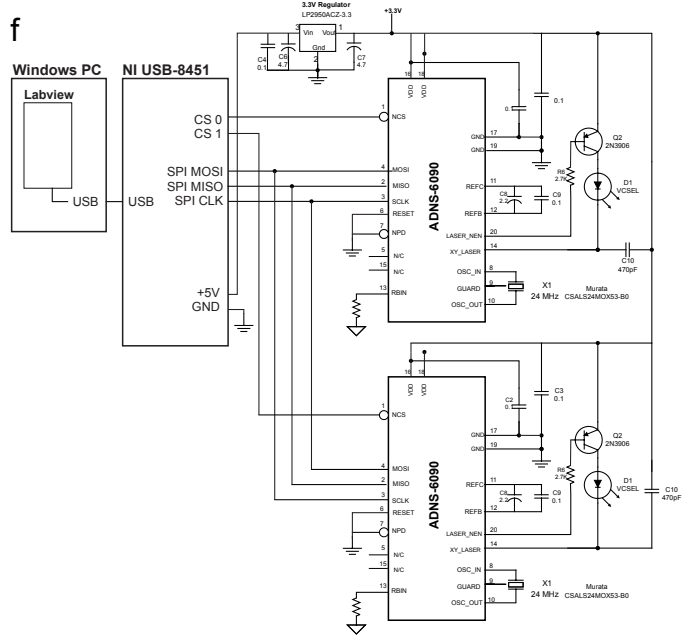
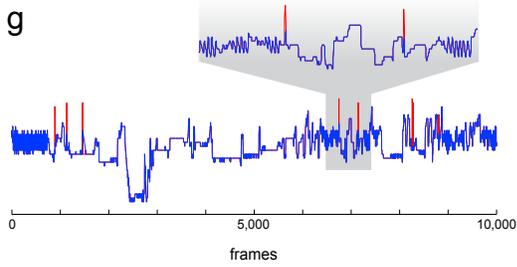
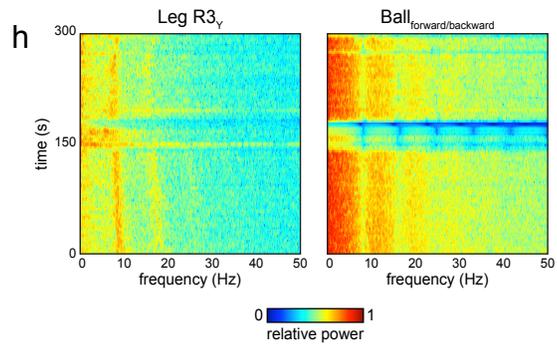

**Supplemental Figure Captions**

**Figure S1|** Additional apparatus details.
(a) Photograph of the leg-tracker setup.
(b) Absorption and emission wavelengths as in Figure 1e, with the optional croconium ball dye included.
(c) Chemical structure of the croconium dye.
(d) Motion of the ball around its axis of rotation (A) is described by a basis set of three component rotational axes (i, j, k), and inferred from the ball's motion at the tracking spots (red).
(e) The appearance of the floating ball in infrared as illuminated by the tracking sensor lasers.
(f) Wiring diagram allowing the Avago ADNS6090 tracking sensors to operate once extracted from their original application in computer mice, as well as communicate with the SPI hub.
(g) A representative vector data trace before (red) and after (blue) clean up using a median filter and removal and interpolated replacement of rare erroneous readings. See Supplemental Note 4.
(h) Sphere motion is sufficiently sensitive to respond to single strides, as shown by time-frequency analysis of an individual leg vector (left) and the forward/backward component of the ball motion (right). Color indicates relative power. The ~9 and ~18Hz leg-motion modes drive corresponding modes in the motion of the ball.

**Supplemental Movie Captions**

**Movie S1|** A fly spontaneously behaving on the leg tracking setup.

Movie depicting the data used by human investigators (left side) and the data used by the *k*-nearest neighbors classifier (right side) for annotating the same 4000 frames of a tethered fly spontaneously behaving upon a floating sphere. Video capture of a fly used for frame-by-frame hand scoring (upper left). Note the infrared fluorescent dye spots upon each leg (orange-pink color). The visibly bright spot on the ball is a reflection of one of the tracking sensors. The 4000 frame-by-frame hand-scoring annotations made by two human investigators (JK and BD) are shown (bottom left). Visualization of the raw instrument data – the 12 leg position vectors and the 3 motion components of the ball (right, first column), and visualization of the higher order features (the derivatives and local standard deviations of the raw data) (right, second and third columns). The annotation made by the classifier using the raw data, the higher order data and low-pass filtering, for each frame (bottom right).

**Movie S2|** The *k*-nearest neighbors classifier annotating a novel behavioral data set.

As in Movie S1, except there is no human hand-scoring, which was done only for training the classifier. Once the KNN has been trained with the manual data, an arbitrary number of additional experiments can be classified with no additional manual scoring required.

## SUPPLEMENTAL NOTES

**Note 1|** Apparatus design details and use.

Previous reports of *Drosophila* spherical treadmills used styrofoam, high-density polyethylene, or polyurethane foam spheres [1-4]. This was not an option for our setup because we needed an optically clear sphere for imaging the dye spots of the leg from below. At first we thought we would need a spherical treadmill that matched the inertial mass of the tiny fruit fly. This led us to try aerogel, a rigid, ultralight and translucent material produced through supercritical drying [5]. However, it was time-consuming and difficult to make uniform spheres of aerogel, the material absorbed infrared and was very hygroscopic.

Next, we tried clear, acrylic spheres (1/4", McMaster-Carr), which were much heavier than a fruit fly, but considerably more spherical than our in-house manufactured aerogel spheres. We were able to drastically reduce the mass of the sphere (down to <7mg) by splitting them in half, hollowing them out using Dremel rotary tools, and then glueing them back together. However, we found that the spheres were too light, and that the fly had trouble running because each stride would displace the sphere instead of spinning it. Therefore, matching the inertial mass of the sphere to the fly was not the correct approach and ultimately, we found the clear, acrylic spheres without modification yielded behaviors that, to our eyes, were the most naturalistic.

We introduced small, scattering surface imperfections on the spheres by lightly rolling them over 400 grit sandpaper to allow the infrared sensors to efficiently track motion. The more scuffed the sphere, the better it was tracked, but at the cost of poorer transmission of the fluorescence from the leg dyes to the cameras below. Thus, we incrementally scuff a sphere and strike a balance between infrared tracking and fluorescence transmission.

Alternatively, the spheres can be coated in the near-infrared absorbing/reflecting croconium dye (Piperdinium, 4-carboxy-1-[5-[3-[5-(4-carboxy-1-piperdinyl0-2-thienyl]-2-hydroxy-4,5-dioxo-2-cyclopenten-1-ylidene]-2(5H)-thienylidene]-, inner salt) to make it visible to the sensors. The croconium dye was prepared using the procedure reported [6]. However, all of the data presented in this work was collected using the sandpaper method.

The housing of the treadmill sphere bearing consisted of the rim of a 150μL Eppendorf tube. During trials, the ball was floated on air from a cylinder of compressed air (Airgas). The air was first bubbled through water to prevent static charge from accumulating, then passed through an adjustable flow regulator, and finally into a plenum to evenly distribute the flow into the air bearing. We used an airflow rate of approximately 440mL per minute.

Flies were grown on standard cornmeal media in 25°C incubators with a 12/12h light-dark cycle. For mounting, a fly was glued on the notum to a stainless steel #0 insect pin (Bioquip) using UV cured adhesive (LED100 system from Electro-Lite Corporation and KOA 300 adhesive from Kemxert Corporation). The pin was glued to a flexible copper wire attached to a male audio jack. The male audio jack-wire-pin module was designed to be small and mobile so that it can be used at the dissecting scope. Once the fly was attached, the module press-fits into a female audio jack that was attached to the micromanipulators (Siskiyou, Inc) of the larger apparatus, which were used for fine adjustments to the animal's position upon the sphere. Animals were allowed to adapt to their positioning in the device for approximately 20-40min, in the dark, prior to data collection.

Infrared sensors (Avago chip# ADNS6090) for tracking the sphere are available in high-performance gaming mice. The wiring diagram of the circuit that allows these sensors to operate once removed from their original circuit boards, as well as communicate with a Serial Peripheral Interface (SPI) hub (National Instruments) is shown in Supplementary Figure S2. See Supplementary Note 2 for details on transforming the four vectors from the two sensors into the three motion vectors of the sphere. Measurement of angular displacement in the three components of rotational motion of the ball was calibrated by manually turning a calibration sphere (rigidly attached to a long, stiff, wire handle) through $\pi$ radians along the turning-in-place and side-stepping dimensions and $\pi/2$ radians along the forward-backward dimension, numerous times and averaging the detected sensor motion.

The dyes were excited by a HeNe laser (632.8 nm, Thorlabs) which passes through a beam expander to ensure the space surrounding the legs was fully illuminated. Since a sphere will act as a lens, a second sphere was positioned within the optics housing to re-collimate the light before reaching the cameras. In conjunction with a beam splitter, two Prosilica GigE cameras (Allied Vision Technologies) were used for imaging the legs from through the spheres. Band-pass filters (Edmund Optics) were used to optimize each camera's detection for either 221ox (HQ650/20m band-pass filter, Chroma) or julox (HQ700/20m band-pass filter, Chroma). Additionally, a short-pass filter (NT49-829, Edmund Optics) was used to block stray light from both the infrared tracking sensors and potential two-photon sources.

In real time computational image processing, the 221ox image was masked out from the julox image within custom LabVIEW interfaces (National Instruments), so that the image from each camera now represented one specific tripod (i.e. a set of three three alternating legs). Leg identity was assigned by simple positional logic. E.g. for the tripod consisting of the front left, middle right and back left legs, the rightmost image dot was assigned the identity of a middle leg, and the frontmost of the remaining two dots a front leg identity.

We used a digital microscope video recorder (ProScope, Bodelin Technologies) for trials where movies were being simultaneously recorded for hand-scoring. We inserted a polyester film filter (Roscolux, medium-blue color filter #83, Roscoe Laboratories) over the microscope to block light from the HeNe laser while still allowing visualization of the sphere, fly, and fluorescing dye spots.

For the optomotor feedback experiments we used a standard 15 inch LCD monitor to display alternating blue (RGB=0000FF, width=4.8cm) and black (RGB=000000, width=2.4cm) colored vertical bars. Optic flow consisted of these vertical bars translating out (10cm/s) from a line of expansion. The monitor was positioned six inches from the tethered fly. The optomotor feedback trial comprised the conditions: 20min in the dark, 20min closed-loop, 20min dark, 20min open-loop, repeat from the beginning once. Closed-loop allowed the turns of the fly to control the position of the line of expansion for the translating vertical bars based on the detected turning-in-place (i) vector. Open-loop disabled the feedback control from the fly and allowed the point of origin for the diverging vertical bars to drift (0.5cm/s) to the point that the entire screen was filled with bars moving either left-to-right or right-to-left. During the 20min open-loop trial, the drift would switch direction every 5min.

**Note 2|** Transforming the tracking sensor data into motion vectors.

The X, Y coordinates from the sensors, integrated over 10 frames in a sliding window fashion as a noise reducing measure, were transformed to generate the 3 motion vectors of the sphere as follows:

- The axis of rotation of the ball $A$ has components whose relative proportions are the proportional extents to which the fly is running forward/backward, sidestepping, or turning in place. Specifically, rotation around $i$ indicates turning in place, $j$ forward/backward walking, and $k$ sidestepping. We need to determine a, b, c such that $A=\{ai, bj, ck\}$ - these are the relative proportions of each type of motion.
- Data from the sensors comes as two vectors $L$ and $R$ each in its own coordinate system (basis): $L=\{x_L, y_L, 0\}$, $R=\{x_R, y_R, 0\}$. First, convert the basis of these vectors to the global ball basis (B=$\{\{i,0,0\},\{0,j,0\},\{0,0,k\}\}$) using a change of basis matrix $C$ derived from the positions of the sensors. Call the converted vectors $L_B$ and $R_B$. These will always be latitude lines with respect to $A$. Therefore $A$ is parallel to a vector which is perpendicular to both $L_B$ and $R_B$. Therefore $A$ is parallel to $A'=L_B \times R_B$.
    - $A'_i$ is the % turning in place
    - $A'_j$ is the % forward/backward translation
    - $A'_k$ is the % sidestepping.
- From $A$, we don't know the speed at which the ball is turning. This is derived from the length of either $L$ or $R$ as a function of their distance from $A$. We calculate it based on whichever sensor is farther from $A$, as this will be a less noisy measurement. Assume, without loss of generality, that $|L|>|R|$, i.e. that $L$ is farther from $A$. The angle between $A$ and $L$ is given by $\cos^{-1}((A \cdot L)/(|A||L|))$. The surface speed of a sphere moving with speed 1 at an angular distance $\theta$ from the axis of rotation is $\sin(\theta)$, thus the speed of rotation of the ball (and the magnitude of the vector whose components are the fly's behavioral components) is
$$M = |L|/\sin(\cos^{-1}((A \cdot L)/(|A||L|)))$$
- While $L_B \times R_B$ is mathematically parallel to $A$, when $R_B$ is nearly parallel to $L_B$, $A'$ will approach 0, and its orientation will become highly sensitive to noise in the measurement of $L$ and $R$. This occurs when $A$ lies in the plane of the bearing housing, i.e. perpendicular to $i$. To the extent that $L_B$ and $R_B$ are parallel, we want to use a different formula than their cross product. Since $L_B$ and $R_B$ are parallel, we can replace their information with their mean, and we can let "the extent to which they are parallel" be the cosine of the angle between $L_B$ and $R_B$:
$$\beta = \cos(\cos^{-1}((L_B \cdot R_B)/(|L_B||R_B|))) = (L_B \cdot R_B)/(|L_B||R_B|)$$
- This is the weighting factor given to the term of our final equation when $L_B$ and $R_B$ are parallel. The term itself is simply the cross product of $i$ with the mean of $L_B$ and $R_B$, which by assumption are both perpendicular to $A$. Thus, the noise-resistant ball-basis axis of rotation is given by:

$$A' = \beta*(L_B+R_B)/2 \text{ X } \{(|L_B|+|R_B|)/2, 0, 0\} + (1-\beta)*L_B \text{ X } R_B$$

**Note 3|** Dyeing the fly legs.

The dye [2.2.1]-oxazine (phenoxazin-5-ium, 3,7-bis(7-azabicyclo[2.2.1]hept-7-yl)-chloride), "221ox," was prepared according to the method described elsewhere [7]. The dye julolidine oxazine (1H,5H,11H,15H-diquinolizino[1,9-bc:1',9'-hi]phenoxazin-4-ium, 2,3,6,7,12,13,16,17-octahydro-, chloride), "julox," was prepared according to the method described elsewhere [8]. Leg-tracking would likely work using quantum dots [9] instead of the dyes, by selecting dots based on their diameters to match the emission spectra of the dyes, but we did not try this approach.

The dyes were dissolved in methyl chloride and then mixed with clear nail polish and spread out into a thin film upon a glass slide to dry. Small pieces were cut out of the dye film with a scalpel. A small dab of UV-cure glue was placed on each leg of the animal under $CO_2$ anesthetization (Fig. 1B, C). A piece of dye film was placed onto the glue drop and then exposed to several seconds ultraviolet light, thus securing the dye to the leg. Dye pieces were placed on the femurs of the fore and hindlegs and the tibias of the middle legs to better separate the dye signals in space for easier tracking. Marked flies were allowed to recover and acclimate overnight.

**Note 4|** Developing the *k*-nearest-neighbors classifier.

We considered feed-forward neural networks and support vector machines (SVM) as non-linear classifiers for our instrument data before settling on *k*-nearest neighbors analysis (KNN). Those alternative methods showed promise, but neural networks are hard to interpret and have path-dependent performance variability. SVM requires computationally intensive optimization of two parameters (in the case of the most generally useful kernel). Importantly, we intend to present our technique to building the classifier as a general approach, and the intuitive underpinnings, computational simplicity, and successful performance of KNN were compelling. The classifier was built as follows.

A training set was built by manually scoring 4000 or 8000 video frames of a fly behaving on the ball while simultaneously recording the 15 raw data vectors. This was done independently by two investigators to assess the variability in manual scoring. Five different flies were scored.

Data were recorded as fast as possible during experiments, resulting in varying inter-frame intervals, with an average recording rate of ~90Hz. Prior to any additional processing, raw data was linearly interpolated to a uniform rate of 100Hz. The videos acquired for training purposed were recorded at 30Hz, and therefore the frames used in conjunction with the manual scores were selected for their correspondence with the time stamps of the video frames. Illumination of the HeNe laser was used as a trigger to synchronize the video and instrument data.

Raw data contained rare frames with erroneous leg-track positions. These typically arose when the dye spots moved out of the HeNe laser excitation, e.g. if they were occluded by the abdomen. Any trial with more than ~1% error frames was discarded. Raw data for the KNN training, and all other experiments, was filtered to remove error positions as follows. A 3-frame median filter was applied to all raw data vectors eliminating single-frame errors. The first frames of errors persisting longer than a single frame were identified by their values deviating from adjacent frames by more than 5 standard deviations of the inter-frame differences. The final frames of these extended errors were identified as the next frame within 5 standard deviations of the value prior to the error, or within 1 standard deviation of the median position across all frames. Values thus flagged as erroneous were replaced with values interpolated linearly from flanking frames. This approach was conservative, but removed essentially all conspicuous errors (Fig. S1g).

To explore which higher-order features were required for sufficient KNN performance, we used cross-validation at the level of flies – i.e. setting aside all of one fly's data, training the KNN classifier with the raw values and higher-order features from the remaining four flies from the training set. Correlation coefficient was used as the KNN metric. Performance was assessed using the withheld fly's raw data and higher-order features. Data from different flies was made comparable by *z*-score normalization within the data from each fly (to eliminate any positioning differences between mountings), then concatenated and *z*-score normalized again (to assure that all training variables would contribute equally to the classifications). The data from the withheld fly was z-score normalized, and then re-normalized using the mean and standard deviation of concatenated training set data.

Performance of the KNN with respect to one set of manual scores was assessed in two ways: the unbalanced accuracy, i.e. the overall percent of classifications matching the manual score, and the "plausible accuracy" which tolerates errors in assignment between behaviors that are only arbitrarily distinguishable. E.g. the precise distinction between standing still and making small postural adjustments is necessarily an arbitrary cutoff. Misclassifications tolerated in this analysis include the above, postural adjustment with the locomotory behaviors, the locomotory behaviors with complex motion, leg1- and head-grooming, and leg3- and abdomen-grooming. We have observed that flies will groom both their hind-legs and abdomen in the same stroke, so this reflects biological overlap in addition to definitional continuity. The plausible accuracy between the manual scoring data sets was 96%, and between the KNN classifier scores and manual (JK) scores 91%. Manual (BD) scores were slightly more difficult for KNN to classify, yielding an 88% plausible accuracy.

Higher-order features characterizing local temporal dynamics in the data were calculated with a sliding window of +/-5frames. Other window sizes were evaluated and did not improve performance. Vector derivatives were calculated as the difference of values from frame *t*-1 to *t* +1. Local cross-correlations were calculated in a pair-wise fashion between all 15 raw values in a sliding window of +/-5frames. Vectors of higher-order features were padded with 0s as necessary. The KNN *k* parameter was optimized to 16 using just the raw data, and re-optimized to 24 once the best training set (raw data + derivatives + local standard deviations) was identified. KNN performance was quite robust to the choice of $4<k<100$. Lastly we noticed that the model was prone to switch between behavioral types more frequently than the manual scorers. Thus a low-pass filter that replaced each frame's classification with the most abundant classification in a +/-5frame sliding window further improved performance.

Raw data collected post-KNN-training was interpolated, filtered for errors, normalized, augmented with higher-order features, and low-pass filtered as above, with the exception that the training samples included data from all 5 training flies, i.e. none were withheld.

**Note 5|** Data analysis.

Ethograms were calculated by populating a 12x12 transition matrix at position $M_{a,b}$ with the number of instances for which a frame of behavior $a$ precedes a frame of behavior $b$. Self-transitions were ignored to focus on inter-behavioral transitions, and because self-transitions dominate the pie chart representations. Rows of the transition matrix were normalized by their totals to yield transition probabilities, which were visualized in a network fashion.

Principle components analysis (PCA) was performed on the (12) percentages of the pie charts and (132=12x11) probabilities of the ethograms, with replicates consisting of the 11 experiments done across 5 different flies over 2 days (or 3 days in the case of one fly). In both cases, all input variables were $z$-score normalized prior to PCA, and the first two principle components were found to explain a minority of the total variance. Statistical significance of the clustering of each fly's day-to-day replicates versus the overall distribution of the trials in Fig 3c and 3e was determined two ways: by $t$-test comparing the mean intra-fly distance in PC-space to the mean inter-fly distance, and by a resampling approach in which the labels on the trial data were shuffled randomly and the intra- and inter-fly distances calculated. The mean and standard deviation of the resampled mean inter-fly distances were compared to the observed distance using a $z$-test. These methods were in very close agreement. The intra-fly trials were not significantly clustered when total percentage of time in each behavior was considered ($p=0.2$, Fig 3c), but were clustered when the transition rates between behaviors were considered ($p<0.001$, Fig 3c).

The average position of each leg was determined across behavioral transitions by identifying target behavioral transitions from the sequence of KNN classified behaviors, and then averaging leg positions in a +/-50frame window flanking the transitions, across all occurrences of the transition.

**SUPPLEMENTARY REFERENCES**